\documentclass[5p,twocolumn,11pt]{elsarticle}

\usepackage{amsmath,amsfonts,amssymb,amsthm, booktabs, tabularx, placeins}

\usepackage{comment}
\usepackage{physics}
\usepackage{graphicx}
\usepackage[mathscr]{eucal}

\usepackage{subfig}

\newcommand{\be}{\begin{equation}}
\newcommand{\ee}{\end{equation}}
\newcommand{\bea}{\begin{eqnarray}}
\newcommand{\eea}{\end{eqnarray}}

\usepackage[dvipsnames]{xcolor}
\usepackage[breaklinks,colorlinks,urlcolor=MidnightBlue,citecolor=WildStrawberry,linkcolor=Fuchsia]{hyperref}
\usepackage{orcidlink}

\begin{document}

\begin{frontmatter}
\title{Fractional attractors in light of the latest ACT observations}

\author[Taltech,KBFI]{Christian Dioguardi\orcidlink{0000-0002-6133-0383}}

\author[NYU,MUN]{Antonio J. Iovino\orcidlink{0000-0002-8531-5962}}

\author[KBFI]{Antonio Racioppi\orcidlink{0000-0003-4825-0941}}

\affiliation[Taltech]{organization={Tallinn University of Technology},
            addressline={Akadeemia tee 23}, 
            city={Tallinn},
            postcode={12618}, 
            country={Estonia}
}
\affiliation[KBFI]{organization={National Institute of Chemical Physics and Biophysics},
            addressline={Rävala 10}, 
            city={Tallinn},
            postcode={10143}, 
            country={Estonia}
}
\affiliation[NYU]{organization={New York University, Abu Dhabi},
            addressline={PO Box 129188 Saadiyat Island}, 
            city={Abu Dhabi},
            country={UAE}
}
\affiliation[MUN]{organization={Institute for Theoretical Physics, University of Münster},
            addressline={Wilhelm-Klemm-Straße 9}, 
            city={Münster},
            postcode={48189}, 
            country={Germany}
}
\begin{abstract}
In light of the latest results from ACT observations we review a class of potentials labeled as fractional attractors, that can originate from Palatini gravity.
We show that, for certain choices of the scalar potential $V(\phi)$, the fractional attractors predict both a spectral index $n_s$ and a tensor-to-scalar ratio $r$ that fall within the $1\sigma$ region of the combined ACT+Planck data for a wide range of parameters. We also provide a numerical fit for the parameter space of this models in the case of a simple quadratic and quartic fractional potential.
\\
\\MS-TP-25-09
\\
\\Emails: christian.dioguardi@kbfi.ee, antoniojunior.iovino@uniroma1.it, antonio.racioppi@kbfi.ee
\end{abstract}


\end{frontmatter}

\section{Introduction} 
\label{sec:Introduction}
The observation of the cosmic microwave background radiation (CMB) provides strong evidence for the cosmological principle. To account for the observed flatness and homogeneity, a period of accelerated expansion, known as inflation, is required in the very early universe \cite{Starobinsky:1980te,Guth:1980zm,Linde:1981mu,Albrecht:1982wi}.

Support for inflationary cosmology has recently been reinforced by the latest data release from the Atacama Cosmology Telescope (ACT) \cite{ACT:2025fju,ACT:2025tim}. However, when combined with the Year 1 data from DESI \cite{DESI:2024uvr,DESI:2024mwx}, these results significantly shift the constraints on the spectral index $n_s$ compared to the previous Planck 2018 dataset \cite{Planck2018:inflation}, disfavoring the commonly studied $R^2$ Starobinsky inflationary model at the $2\sigma$ level \cite{Starobinsky:1980te}.

Indeed, while the Planck 2018 analysis previously determined $n_s = 0.9651 \pm 0.0044$ \cite{Planck2018:inflation,Planck:2018vyg}, the latest constraints, obtained from a joint analysis of Planck and ACT data, yield $n_s = 0.9709 \pm 0.0038$. Furthermore, when Planck, ACT, and DESI data (ACT+Planck) are combined, the constraint tightens to $n_s = 0.9743 \pm 0.0034$ \cite{ACT:2025tim}, deviating from the original Planck result by approximately $2\sigma$.

Even though too early to make final claims, the current status is that 
the commonly studied Starobinsky model is now disfavored at the $2\sigma$ level, but many other inflationary models remain compatible with the latest ACT data release. For instance, these includes
various setups of non-minimal Coleman-Weinberg inflation \cite{Marzola:2016xgb,Jarv:2017azx,Racioppi:2018zoy,Kannike:2018zwn,Racioppi:2019jsp,Gialamas:2020snr,Gialamas:2021rpr,Racioppi:2021ynx,Kannike:2023kzt} and of non-minimal metric-affine gravity \cite{Gialamas:2024uar,Racioppi:2024pno,Racioppi:2024zva,Bostan:2025vkt}, generalized hilltop models \cite{Lillepalu:2022knx}, 
polynomial inflation with non-minimal coupling to gravity \cite{Bezrukov:2007ep,Bezrukov:2014ipa,Kallosh:2013tua,Rubio:2018ogq}, pole inflation and KKLTI inflation \cite{Galante:2014ifa,Kallosh:2019hzo}, polynomial and hybrid $\alpha$-attractors \cite{Kallosh:2022feu,Kallosh:2022ggf,Kallosh:2019jnl,Dalianis:2018frf}, supersymmetric hybrid
inflation\,\cite{Buchmuller:2014epa,Schmitz:2018nhb} and the double inflection point polynomial potential \cite{Allegrini:2024ooy}. In particular, the results from ACT attracted the attention from the community and new models to fit the data have been recently proposed \cite{Kallosh:2025rni,Aoki:2025wld}. 

Among the others, the Palatini formulation of gravity \cite{Koivisto:2005yc,Bauer:2008zj,Gialamas:2023flv} (and refs. therein) also provides a promising framework to understand inflation.
In the Palatini formalism the affine connection is a priori independent from the metric. When the gravity sector is non-minimally coupled to matter fields the phenomenological predictions differ from the prediction of metric gravity. In particular, the Palatini framework was shown to have many appealing features to build inflationary setups. An example is that of $F(R)$ models, for which the gravity sector is taken to be a general function of the Ricci scalar. This class of models generates asymptotically flat potentials that can be used to describe experimentally viable slow-roll inflation (e.g. \cite{Enckell:2018hmo,Dioguardi:2021fmr,Dioguardi:2022oqu} and refs. therein). 
A generalization of the $F(R)$ theories namely $F(R,X)$ can provide a class of fractional attractor potentials \cite{Dioguardi_2024,Dimopoulos:2025fuq}, which exhibits an attractor behavior that generalizes the polynomial $\alpha$-attractors \cite{Kallosh:2022feu}. 

In this letter we prove in a model-independent way that the class of fractional attractor potentials fits the recent combined observations from ACT+Planck for a wide range of parameters.

The letter is organized as follows: in section \ref{sec:fractional:attractors} we introduce the model based on fractional attractor potentials and provide the inflationary predictions in the strong-coupling regime. In section \ref{sec:ACT:fit} we fit the two most simple examples, i.e. a quadratic and quartic fractional potential to the $2\sigma$ region of the Planck-ACT data and then, we conclude in section \ref{sec:conclusions}.


\section{Fractional attractor potentials} \label{sec:fractional:attractors}
Consider the action
\be
S = \int d^4x \sqrt{-g^E}\qty(\frac{m_P^2}{2}R-\frac{1}{2}\partial_\mu\phi \partial_\nu \phi-U(\phi)), \label{eq:action:frac}
\ee
with $U(\phi)$ the inflaton potential. It is well known that data \cite{ACT:2025tim,Planck2018:inflation} strongly favors nearly flat inflaton potentials. One of the simplest way to construct an asymptotically flat potential is via the parametrization
\be\label{eq:fractional}
U(\phi) = V(\phi)\left(1+8\alpha \frac{V(\phi)}{m^4_P}\right)^{-1},
\ee
where $\alpha$ is a dimensionless parameter and $V(\phi)$ a generating function with the dimensions of a scalar potential.
When $\alpha \to \infty$, the potential $U$ becomes asymptotically flat with $U(\phi)\approx \frac{m^4_P}{8\alpha}$ for any specific choice of $V(\phi)$.
The potential in eq. \eqref{eq:fractional} is a generalization of the polynomial $\alpha$-attractors studied in \cite{Kallosh:2022feu}.  Such a setup can be built in the context of Palatini gravity \cite{Dioguardi_2024} (see\,\ref{app}).

The CMB observables for the potential in Eq. \eqref{eq:fractional} can be computed introducing the slow-roll parameters:
\bea
\epsilon  (\phi) &=& \frac{m_P^2}{2}\qty(\frac{U'(\phi)}{U(\phi)})^2 \, , \label{eq:epsilon}
\\
\eta  (\phi) &=& m_P^2\frac{U''(\phi)}{U(\phi)} \, . 
\eea
The number of $e$-folds of Universe expansion can be computed directly from $\epsilon$:
\be
N =  \frac{1}{m_P^2}\int_{\phi_{\textrm{end}}}^{\phi_N} {\rm d}\phi \, \frac{U(\phi)}{U'(\phi)} ,
\label{eq:Ne}
\ee
where $\phi_{end}$ is obtained by imposing the exit condition from slow-roll inflation\footnote{The condition $|\eta|\simeq1$ can also trigger the end of the slow-roll approximation. We explicitly checked that during the inflationary evolution, i.e. until $\epsilon = 1$ , the condition $|\eta|\ll1$ was satisfied for the whole parameter space considered.}, i.e. $\epsilon(\phi) = 1$. The field value $\phi_N$ at the time a given scale left the horizon is given by fixing $N$.
From this, one can straightforwardly compute the CMB observables,
i.e. the spectral index $n_\textrm{s}$ and the tensor-to-scalar ratio $r$:
\bea
n_\textrm{s}  &=& 1+2\eta  (\phi_N)-6\epsilon  (\phi_N) \, ,  \label{eq:ns} \\
r  &=& 16\epsilon  (\phi_N) \,  , \label{eq:r}
\eea
and the amplitude of the scalar power spectrum\,\cite{Planck2018:inflation}: 
\be
 A _\textrm{s} = \frac{1}{24 \pi^2 m^4_P }\frac{U(\phi_N)}{\epsilon  (\phi_N)} \simeq 2.1 \times 10^{-9} \, .
 \label{eq:As:th}
\ee
We can find a model independent expression for $r$ in the strong coupling limit $\alpha \to \infty$. As mentioned before, the potential asymptotically approaches the plateau $U(\phi)\rightarrow \frac{m^4_P}{8\alpha}$ for any specific choice of $V(\phi)$. Therefore, inserting this value in the numerator of Eq.\eqref{eq:As:th} and using Eq.\eqref{eq:r}, it is straightforward to show that in a model-independent way
\be
 r  \approx \frac{1}{12 \pi^2 A_s \alpha} \, . \label{eq:r:limit}
\ee
This means that we can arbitrarily lower $r$ by increasing $\alpha$, in the same way of the quadratic Palatini models studied in \cite{Enckell:2018hmo}. 
%

On the other hand, the predictions for $n_s$ strongly depend on the choice of $V(\phi)$. As reference cases, we consider potentials in the following monomial form:
\be
V(\phi) = \lambda_k \phi^k, \qquad \lambda_k = \frac{\lambda^k}{k!} \label{eq:V:k}
\ee
where the unusual prefactor, $\lambda_k$, is chosen for numerical convenience and we implicitly performed the redefinition $\phi \to \phi + \phi_N$, which does not affect the kinetic term of the inflaton, but simplifies its potential.
For completeness, it is trivial to show that we can extend the validity of our results also to models in which $0<k<1$.
\begin{figure}[!t]
	\centering
\includegraphics[width=0.495\textwidth]{ 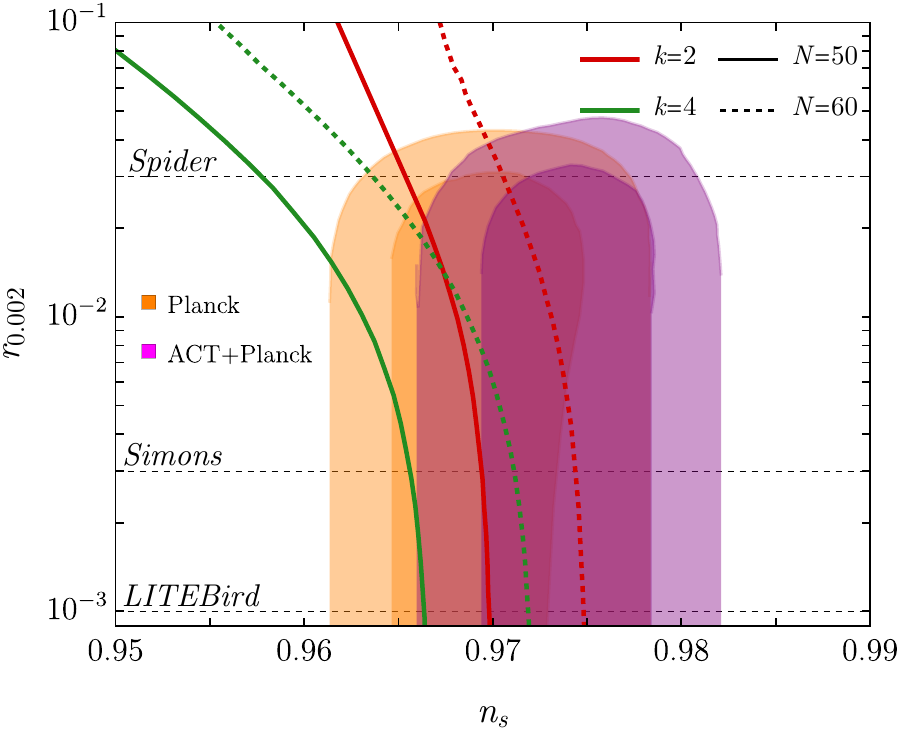}
	\caption{ \it 
Constraints in the $r$ vs. $n_s$ plane using the analysis of Planck data\,\cite{Planck2018:inflation} (orange regions) and the ACT+Planck\,\cite{ACT:2025tim} (magenta regions). The plot includes 68\% and 95\% confidence contours.
The red and green lines represents respectively the predictions of the models increasing $\alpha$ for the benchmark cases $k=2$ and $k=4$ in the potential of Eq.\,\ref{eq:V:k}. Solid and dashed lines are inflationary predicitons with respectively $N=50$ and $N=60$ $e$-folds. In addition, we plot the predicted constraints on $r$ from future experiments such as Spider \cite{SPIDER:2017xxz}, Simons Observatory \cite{SimonsObservatory:2018koc}, and LiteBIRD \cite{Matsumura:2013aja}.
 }
\label{fig:ModelsCMBConstraint}
\end{figure}

In general, we have:
\bea
N &=& \frac{\phi ^2 }{2m_P^2 k }\left(\frac{16 \alpha  \lambda_k  }{(k+2) m_P^4}\phi ^k+1\right)\Bigg|^{\phi^{end}}_{\phi_N},
\\
r  &=& 16 \epsilon_k (\phi_N),
\\
n_s &=& 1 - 6\epsilon_k (\phi_N) + 2\eta_k (\phi_N),
\\
A_s &=& \frac{\lambda_k  \phi ^{k+2}}{12 \pi ^2 k^2 m_P^6}\left(\frac{8 \alpha  \lambda_k}{m_P^4}  \phi ^k+1\right),
\eea
where 
\bea
 &&\hspace*{-1cm} \epsilon_k(\phi) = \frac{m_P^2 k^2}{2 \phi ^2 }\left( \frac{8\alpha \lambda_k}{m_P^4}\phi^k+1\right)^{-2},\\
&&\hspace*{-1cm} \eta_k(\phi) = \frac{m_P^2 k  \left(k-1-\frac{8 \alpha  (k+1) \lambda_k}{m_P^4}  \phi ^k\right)}{\phi ^2 \left(\frac{8 \alpha  \lambda_k}{m_P^4}  \phi ^k+1\right)^2}.
\eea
In the strong coupling limit $\alpha \rightarrow \infty$, we get
\bea
r  &\sim& 0,
\\
n_s  &=& 1-\frac{k+1}{k+2}\frac{2}{N}.\label{eq:ns:plus}
\eea
\begin{figure}[!t]
	\centering
\includegraphics[width=0.495\textwidth]{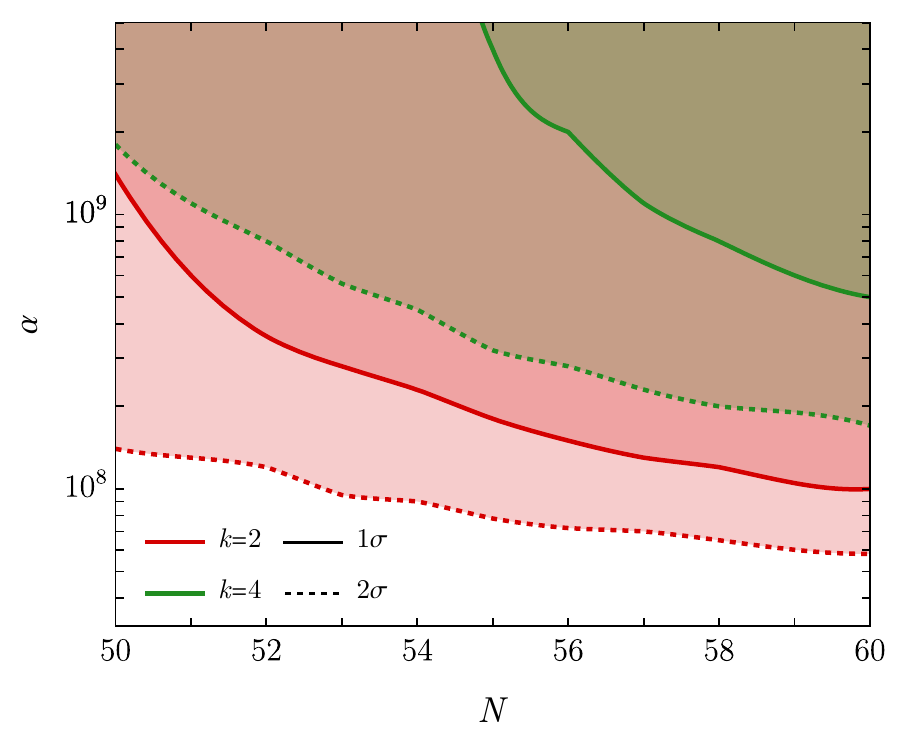}
	\caption{ \it 
Minimum values of $\alpha$ as function of the number of $e$-folds $N$ in order to have inflationary parameters $(r,n_s)$ compatible at $1\sigma$ (solid lines) and $2\sigma$ (dashed lines) confidence levels with the recent ACT+Planck dataset\,\cite{ACT:2025tim} for the two benchmark potentials $k=2$ (red) and $k=4$ (green).
 }
\label{fig:alpha}
\end{figure}

This implies that in this limit, considering the range $0<k<\infty$, the value for $n_s$ is constrained between:
\begin{align}\label{eq:ns:bounds}
0.96\lesssim& n_s\lesssim0.98,& \text{at $N$ = 50},\\
0.967\lesssim& n_s\lesssim 0.983,& \text{at $N$ = 60}.
\end{align}
\section{Fitting the latest ACT release}\label{sec:ACT:fit}
In order to give a more specific fit for the parameter space, we analyze the simple cases with $V(\phi)$ given by Eq.\eqref{eq:V:k} for the choices $k=2$ and $k=4$.

In Fig.\,\ref{fig:ModelsCMBConstraint} we report the inflationary parameters $(r,n_s)$ for the two benchmark potentials. We vary the values of $\alpha$ for $N=50$ and $N=60$ and fix $\lambda_k$ by imposing the observed value for the scalar perturbations $A_s \sim 2.1 \cdot 10^{-9}$. In Fig.\,\ref{fig:alpha} we show the minimum values of $\alpha$ as function of the number of $e$-folds $N$ in order to have inflationary parameters $(r,n_s)$ compatible at $1\sigma$ and $2\sigma$ confidence levels with the recent ACT+Planck dataset\,\cite{ACT:2025tim} for the same benchmark potential. 

In agreement with the analytic asymptotic estimation in Eq.\eqref{eq:ns:plus} we can see from our plots that, for the quadratic model, i.e. $k=2$, there is always a value of $\alpha$ for which the inflationary parameters lie inside the confidence level $1\sigma$ for the new dataset, and only for $N\gtrsim70$ the model asymptotically lies outside the $1\sigma$ region. The situation for the quartic model, i.e $k=4$, is different. Indeed, while for the old Planck dataset the model was in agreement at $1\sigma$ with $N=50$, in the quartic case only $N\gtrsim55$  has inflationary parameters compatible at $1\sigma$ with the new dataset.

\section{Conclusions}\label{sec:conclusions}

After the combined analysis of ACT+Planck data the Starobinsky Model is currently disfavored at $2\sigma$ level. For this reason we provided a simple effective potential that can satisfy the new bounds on the spectral index $n_s = 0.9743\pm 0.0034$ in a model independent way for a wide choice of the free parameters. The effective potential is a straightforward generalization of the polynomial $\alpha-$attractors that can be obtained in the context of Palatini gravity.
We also explicitly showed the parameter space compatible with the new dataset for the case of a quadratic and quartic monomial potential, showing how both potentials predict CMB observables inside the $2\sigma$ region in the range $50\leq N \leq60$, with the quadratic model being favored.

\appendix

\section{Palatini F(R,X)}\label{app}

In this appendix we briefly describe a possible setup to generate a potential in the form of fractional attractors\footnote{This setup was first introduced in \cite{Dioguardi_2024} as a generalization of the Palatini $F(R)$ models, in order to solve inflationary issues beyond slow-roll for models containing $R^n$ terms with $n>2$.}.
Consider the action
\be
S = \int d^4x \sqrt{-g^J}\qty(\frac{F(R_X)}{2} - V(\phi)),\label{eq:action:FRX}
\ee
where $V(\phi)$ is the inflaton scalar potential and $F(R_X)$ is an arbitrary function of its argument. We define $R_X =  R_J+m_P^{-2} X$ with $X = -g^{\mu\nu}_J \partial_\mu \phi \partial_\nu \phi$ denoting the inflaton kinetic term and $R_J = g^{\mu\nu}_J R_{\mu\nu}(\Gamma)$ where $R_{\mu\nu}(\Gamma)$ is the Ricci tensor built from the Palatini formulation. 
Introducing an auxiliary field $\zeta$ as in \cite{Dioguardi_2024}, the action becomes:
\be
S = \int d^4x\sqrt{-g^J}\qty(\frac{F(\zeta) + F'(\zeta) (R_X -\zeta)}{2} - V(\phi)) \label{eq:action:ST} \, ,
\ee
where the symbol $'$ indicates differentiation with respect to the argument of the function. It is easy to check that action \eqref{eq:action:FRX} is obtained from action \eqref{eq:action:ST} by taking the solution of the equation of motion for $\zeta$ i.e. $\zeta = R_X$.
In order to obtain a theory linear in $R$ we can move to the Einstein frame with the help of a conformal transformation $g_{\mu\nu}^E =m_P^{-2} F'(\zeta) g_{\mu\nu}^J$. With such transformation the theory is minimally coupled to the metric $g_{\mu\nu}^E$ and the action reads
\be \label{Einstein_action}
S = \int d^4x \sqrt{-g^E} \qty(\frac{m_P^2}{2}R_E - \frac{1}{2}g^{\mu\nu}_E \partial_\mu \phi \partial_\nu \phi - U(\zeta,\phi)) \, ,
\ee
with
\be \label{eq:einstein_potential}
U(\zeta,\phi) = \frac{m_P^4 V(\phi)}{F'(\zeta)^2} -\frac{m_P^4F(\zeta)}{2 F'(\zeta)^2} + \frac{m_P^4\zeta}{2 F'(\zeta)} \, .
\ee
By restricting our choice to the case:
\be
F(R_X) = m_P^2 R_X + \alpha {R_X}^2 \, , \label{eq:FRX:2}
\ee
and computing the equation of motion of $\zeta$ we can immediately find the solution
\be
 \zeta= \frac{4 V(\phi)}{m_P^2},
\ee
and hence:
\be\label{eq:Pot1}
U(\phi) = \frac{V(\phi)}{1+8\alpha \frac{V(\phi)}{m_P^4}}.
\ee

which gives exactly the fractional potential \eqref{eq:fractional}. For more details on the $F(R_X)$ setup, see \cite{Dioguardi_2024}.
To conclude, we again stress that, even though similar in form, our construction is more general than the one in \cite{Kallosh:2022feu}.

\section*{Acknowledgments}
We thank K. Schmitz for constructive discussions and comments on the draft.
The work of C.D. and A.R. was supported by the Estonian Research Council grants PRG1055,  RVTT3, RVTT7 and the CoE program TK202 ``Foundations of the Universe'’.  This article is based upon work from COST Action
CosmoVerse CA21136, supported by COST (European Cooperation in Science and Technology).
A.J.I. thanks the University of Münster for the kind hospitality during the realization of this project.

\bibliographystyle{elsarticle-num} 
\bibliography{main}

\end{document}